\documentclass[twocolumn,showpacs,preprintnumbers,superscriptaddress,amsmath,amssymb]{revtex4}
\usepackage{graphicx}
\usepackage{dcolumn}
\usepackage{bm}
\usepackage{color}

\begin{document}
\title{Landau Damping of Baryon Structure Formation in the Post Reionization Epoch}
\author{Feng-Yin Chang}%
\email{fychang@ntu.edu.tw}
\affiliation{Leung Center for Cosmology and Particle Astrophysics, National Taiwan University, Taipei 106, Taiwan.}
\affiliation{Institute for
Astrophysics, National Taiwan University, Taipei 106, Taiwan.}
\author{Pisin Chen}
\email{pisinchen@phys.ntu.edu.tw} 
\affiliation{Leung Center for Cosmology and Particle Astrophysics, National Taiwan University, Taipei 106, Taiwan.}
\affiliation{Institute for
Astrophysics, National Taiwan University, Taipei 106, Taiwan.}
\affiliation{Kavli Institute for
Particle Astrophysics and Cosmology, SLAC National Accelerator
Laboratory, Menlo Park, CA 94025, USA.}

\begin{abstract}
It has been suggested by Chen and Lai that the proper description of the large scale structure formation of the universe in the post-reionization era, which is conventionally characterized via gas hydrodynamics, should include the plasma collective effects in the formulation. Specifically, it is the combined pressure from the baryon thermal motions and the residual long-range electrostatic potentials resulted from the imperfect Debye shielding, that fights against the gravitational collapse. As a result, at small-scales the baryons would oscillate at the ion-acoustic, instead of the conventional neutral acoustic, frequency. In this paper we extend and improve the Chen-Lai formulation with the attention to the Landau damping of the ion-acoustic oscillations. Since $T_e\approx T_i$ in the post-reionization era, the ion acoustic oscillations would inevitably suffer the Landau damping which severely suppresses the baryon density spectrum in the regimes of intermediate and high wavenumber $k$. To describe this Landau-damping phenomenon more appropriately, we find it necessary to modify the filtering wavenumber $k_f$ in our analysis. It would be interesting if our predicted Landau damping of the ion-acoustic oscillations can be observed at high redshifts.
 
\end{abstract}
\maketitle

\section{Introduction}



The evolution of the cosmic structure has been an important subject in cosmology. In the post-decoupling era the neutral atoms (the baryons) ceased to feel the pressure from the photons. So the only force that fought against the gravitational pull was its own thermal pressure. Due to the Hubble expansion, the baryon temperature gradually decreased and so was its thermal pressure. The situation changed drastically at the subsequent reionization epoch, during which baryons turned into fully ionized plasmas and were heated up to $T\sim 10^4 K$ by the photo-ionization process \cite{Barkana}. Such reheating process raised the Jeans scale to the size of the galaxies \cite{Mj} and as a result the course of the structure formation was significantly altered. Traditionally the study of the baryon evolution has been conducted by staying within the gas hydrodynamics framework while changing the mean molecular weight between the pre- and the post-reionization epochs to account for the effects of ionization. However, it is known that gas hydrodynamics, which assumes local charge neutrality, cannot exhibit crucial collective phenomena that is unique to plasma, such as Landau damping. It is therefore natural to wonder whether relevant plasma collective effects in cosmic evolution might have been overlooked. 

 In 2007, one of us (Chen) and Lai first considered the plasma effects on the baryon structure formation in the post-reionization epoch, where the conventional baryon acoustic waves at small scales are replaced by the ion acoustic waves in plasmas \cite{plasma suppression}. A new formulation based on the Maxwell-Einstein-Boltzmann equations for pure hydrogen ions was introduced in that paper. To obtain the baryon power spectrum, the authors employed the instantaneous Jeans wavenumber $k_J$ and interpolated the solutions from the opposite limits of the wavenumber $k$. Interesting as it is, this new plasma hydrodynamic formulation has some shortcomings, however. In particular, the spectrum in the intermediate regime of $k$, which is the most crucial regime of interest, was unresolved. To properly account for the impact of the Hubble expansion, it was suggested that the relevant scale should be the integrated and smoothed `filtering wavenumber' $k_f$ rather than the instantaneous Jeans wavenumber $k_J$ \cite{filtering,sound:Gnedin}. Motivated by the reasons, in this paper we analytically re-solve the baryon density spectrum under the Landau damping from the corrected plasma hydrodynamic equation of hydrogen-helium plasmas. Having the solved spectrum, we are able to investigate the impacts of the Landau damping effect on the evolution of baryon overdensity and on the filtering wavenumber $k_f$ due to the decaying amplitude of baryon oscillation modes.
 
 The layout of the paper is the following: In Sec.\ref{ion} we briefly review the concept of ion acoustic wave and its associated new sound speed for baryons. In Sec.\ref{spectrum} we introduce the Landau damping of the baryon density spectrum. This is carried out through the analytical solution of the plasma hydrodynamics equation. Then in Sec.\ref{res} we investigate the influence of Landau damping on the baryon spectrum and derive a modified filtering wavenumber $\tilde{k}_f$ in place of $k_f$. Finally we compare the spectra obtained from plasma and gas approaches, respectively, and discuss the possible ways to distinguish the two models.

\section{Ion Acoustic Formulation}
\label{ion}
  An ordinary sound wave in a neutral gas is a pressure wave that propagates through the thermal collisions among neighboring particles. Without collisions, the sound wave would not occur. However for an ion acoustic wave in a fully ionized plasma, an extra repulsive, non-collisional, electrostatic pressure exists due to the imperfect Debye shielding at finite temperature. Hence an ion acoustic wave can travel through the medium without collisions \cite{Chen,Krall}. Based on the Chen-Lai formulation \cite{plasma suppression,Lai} and the fluid equations for matter overdensites \cite{filtering,eqn:Shapiro}, we obtain the coupled fluid equations for hydrogen-helium baryon (ignoring the electron mass):
\begin{subequations}
\begin{eqnarray}
 \ddot{\delta}_{dk}+2H \dot{\delta}_{dk}
    =4\pi G \bar{\rho} \left(f_b\delta_{bk}+f_d\delta_{dk}\right),\label{dmh}\\
\ddot{\delta}_b+2H \dot{\delta}_b+
\frac{k^2}{a^2}\tilde{c}_s^2\delta_b
    = 4\pi G \bar{\rho}\left(f_b\delta_{bk}+f_d\delta_{dk}\right),\label{baryon}
\end{eqnarray}
\end{subequations}
where the subscripts $b$ and $d$ indicate the baryon and the dark matter, respectively, $\delta\equiv \delta\rho/\rho$ is the fractional overdensity, $f_b\equiv \Omega_b/\Omega_m$ and $f_d\equiv \Omega_d/\Omega_m $ are the baryon and dark matter fractions and $\tilde{c}_s$ is the sound speed for ionic baryons. According to plasma physics, the sound speed is defined by 
\begin{eqnarray*} 
 \tilde{c}_s^2= \delta P_b/\delta \rho_b+\delta \Phi_{em}/\delta \rho_b,
\end{eqnarray*}
where
 \begin{subequations}
\begin{eqnarray}
      \delta \rho_b &=& m_H\delta n_H+m_{He}\delta n_{He},\label{rhob}\\
     \delta P_b&=&\gamma_b k_BT(\delta n_H+\delta n_{He}),\label{pb}\\
      \delta \Phi_{em} 
             &=&\gamma_e k_BT(\delta n_H+Z\delta n_{He}),\label{phi}
\end{eqnarray}
\end{subequations}
 $Ze$ is the helium ion charge and $\Phi_{em}$ is the combined residual electrostatic potential of hydrogen and helium. Here we have assumed the temperature $T_e\thickapprox T_H \thickapprox T_{He}\thickapprox T$ because the time scale for H-He plasma to reach thermal equilibrium is short compared with that of the duration of the reionization epoch \cite{teq}. In a typical plasma $\gamma_e=1$ for isothermal compression condition is often assumed since the thermal velocities of electrons were $\sim \sqrt{m_p/m_e}$ times larger than the fluid sound speed, which enables electrons to interact with each other over several wavelengths within one oscillation \cite{Krall}. Hence invoking the primordial helium abundance $Y\sim 0.24$ \cite{Yp} deduced from the big bang nucleosynthesis (BBN), we can express the ion sound speed $\tilde{c}_s$ in terms of the effective mean molecular weight $\tilde{\mu}$, which combines  
$\gamma_e$ and $\gamma_b$, 
\begin{equation}
 \tilde{c}_s=\sqrt{\frac{5}{3}\frac{k_B T}{\tilde{\mu} m_p}}, 
\end{equation} 
with $\tilde{\mu}=0.75$ for $Z=2$ and $\tilde{\mu}=0.77$ for $Z=1$. Note that in contrast $\mu=0.59$ and $0.62$ for $Z=2$ and $1$, respectively, for the neutral gas.  In general, the ion fluid leads to a lower sound speed by a factor of $\sqrt{\tilde{\mu}/\mu}\sim 1.12$, which corresponds to $\sim 10\%$ increase of the Jeans wavenumber $k_J$.
 
\section{Baryon Density Spectrum with Landau Damping} 
\label{spectrum}

 It is known that plasma waves suffer the Landau damping when the thermal velocity of the plasma particles is slightly below the phase velocity velocity. Residing on the down-slope of the Maxwellian particle velocity distribution, the plasma wave tends to lose energy to more particles and gain energy from less particles. This results in the decline of the plasma wave amplitude. Learning from that, we expect the similar effect on the small-scale baryon oscillations, the ion acoustic modes. Mathematically, Landau damping is a consequence of the singularity $\omega-\mathbf{k}\cdot \mathbf{v}$ in the Vlasov equation, which is absent in the fluid theory. To account for the Landau damping effect in the fluid model, one may separately calculate the Landau damping rate and then multiply it to the imaginary part of the plasma fluid solution.
 
 To analytically study the baryon density spectrum under the Landau damping effect, we start with the coupled fluid equations, Eq.~(\ref{dmh}) and Eq.~(\ref{baryon}), with $f_b=0$ and $f_d=1$ for simplicity.
  Replacing the variable $t$ by $a$ via the relation $t\propto a^{3/2}$ in the matter-dominant era, we arrive at
  \begin{subequations}
  \begin{eqnarray}
  & &\frac{\partial^2 \delta_{dk}}{\partial a^2}
  +\frac{3}{2a}\frac{\partial \delta_{dk}}{\partial a}
  = \frac{3}{2a^2}\delta_{dk}\label{dma},\\
  & & \frac{\partial^2 \delta_{bk}}{\partial a^2}
  +\frac{3}{2a}\frac{\partial \delta_{bk}}{\partial a}+
   \frac{3}{2a^2}\frac{k^2}{\tilde{k}_J^2}\delta_{bk}
   =\frac{3}{2a^2}\delta_{dk},\label{baryona}
  \end{eqnarray} 
  \end{subequations}
where $\tilde{k}_J$ is the co-moving Jeans wavenumber of the ionic baryon. The solution of Eq. (\ref{dma}) gives $\delta_{dk} \propto D_{+}(a)\propto a$ when dropping the decaying mode, while the solution of Eq.~(\ref{baryona}) depends on the thermal history of reionization. We adopt the temperature evolution of the intergalactic medium (IGM), $T\varpropto a^{-0.88}$, which results from the balance between the photo-heating process and the Hubble expansion cooling process \cite{Rees}. In fact the recent measurements indicate a second reheating process at $z\sim 3$ due to the ionization of He$_{\mathrm{II}}$ \cite{IGM_T1,IGM_T2}. To simplify the case, we assume a single reheating process and the temperature evolution $T=T_0a^{-1}$ which leads a constant Jeans wavenumber $\tilde{k}_J(a)=\tilde{k}_{Jion}$, where the subscript $ion$ stands for the ending time of the reionization process. The equation with arbitrary thermal evolution can be analytically solved via Green's function method \cite{solution:Matarrese}.
  
  Inserting the solution of Eq.~(\ref{dma}), Eq.~(\ref{baryona}) can be rewritten as
\begin{equation}
      \frac{\partial^2 \Delta_{bk}}{\partial \alpha^2}+
     \frac{3}{2\alpha} \frac{\partial \Delta_{bk}}
     {\partial \alpha}
     + \frac{3}{2\alpha^2}\kappa^2 \Delta_{bk}
     =\frac{3}{2\alpha}\label{baryon_mod1}
\end{equation}  
 with $\alpha \equiv a/a_{ion}$, $\Delta\equiv \delta/\delta_{ion}$ and $\kappa\equiv k/\tilde{k}_{Jion}$. Note that $\Delta_{dk}=\alpha$. Since we are interested in the relation between the baryon and the dark matter evolutions, it is convenient to express the baryon density in terms of that of the dark matter. Using the initial conditions: 1) $\Delta_{bkion}
=\Delta_{dkion}=1$ and 2) $\partial \Delta_{bk}/\partial \alpha=
\partial\Delta_{dk}/\partial \alpha=1$ at $a_{ion}$, we obtain the normalized baryon density spectrum:
\begin{eqnarray}
  \frac{\Delta_{bk}}{\Delta_{dk}}(\alpha,\kappa)
  =\alpha^{-5/4} [A_1 \alpha^{Q/4}+ A_2 \alpha^{-Q/4}]+\frac{1}{1+\kappa^2}, \label{sol}
  \end{eqnarray} 
 where $Q\equiv\sqrt{1-24\kappa^2}$ and
\begin{eqnarray*}
A_1=\frac{5+Q}{2Q}\frac{\kappa^2}{1+\kappa^2}; \qquad A_2=\frac{-5+Q}{2Q}\frac{\kappa^2}{1+\kappa^2}.
\end{eqnarray*}

 As $\kappa^2 >24$, $Q$ becomes imaginary. We may therefore rewrite Eq. (\ref{sol}) as
\begin{eqnarray}
  & &\frac{\Delta_{bk}}{\Delta_{dk}}(\alpha,\kappa^2>1/24)=\nonumber\\
  & &\alpha^{-5/4}\left[\frac{5}{q}\sin\left(\frac{q}{4}\ln \alpha\right) +\cos\left(\frac{q}{4}\ln \alpha\right)\right]+\frac{1}{1+\kappa^2}\nonumber\\
  & &\equiv b_i(\alpha,\kappa)+\frac{1}{1+\kappa^2}, \label{baryq}
\end{eqnarray}
 where $q\equiv \sqrt{24k^2-1}$ and $b_i(\alpha,\kappa)$ denotes the oscillatory part of $\Delta_{bk}/\Delta_{dk}$. If $\kappa \to \infty$ or equivalently $q \to \infty$, $\Delta_{bk}/\Delta_{dk}$ asymptotically approaches $\alpha^{-5/4}\cos(q/4)$. We see the amplitude of the baryon oscillations decreases as $\alpha^{-5/4}$.  

 Because $T_e \approx T_i$, the thermal velocity of the ions is close to the sound speed. We then expect that the Landau damping would occur to the baryon fluid oscillations. By invoking the nonlinear empirical formula which relates the real and the imaginary parts of the ion acoustic oscillation frequency \cite{Chen}, we introduce the Landau damping effect to Eq.~(\ref{baryq}) by multiplying the oscillation term with an exponential factor, 
\begin{equation}
 \frac{\Delta_{bk}}{\Delta_{dk}}(\alpha,\kappa^2>1/24)=e^{-(q\omega_i/4\omega_r)\ln\alpha}b_i(\alpha,\kappa)+\frac{1}{1+\kappa^2},\label{baryld}
\end{equation} 
where
\begin{equation}
   \frac{\omega_i}{\omega_r}\cong 1.1 \left(\frac{T_e}{T_b}
   \right)^{7/4} e^{-(T_e/T_b)^2}\approx 0.4.
\end{equation}
This means the ion acoustic wave would be Landau-damped to one e-fold of its initial amplitude within just $\sim 0.4$ of one oscillation period. Such severe damping effect causes the baryon oscillations to be totally smoothed out at late times.
\begin{figure}[tb]
\begin{center}
$\begin{array}{c}
\includegraphics[width=8cm]{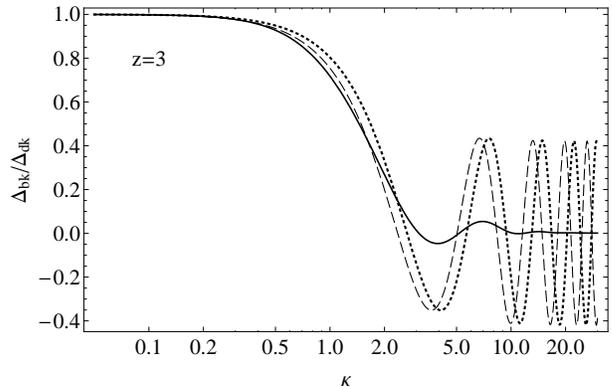}
\end{array}$
\end{center}
\caption{ The baryon density spectrum under Landau damping(solid) at $z=3$ with $z_{ion}=7$, superimposed by the dashed and dotted curves standing for the results from gas hydrodynamics ($\mu$) and that from plasma physics before applying landau damping ($\tilde{\mu}$) respectively.
}\label{solo}
\end{figure}
\begin{figure}[tb]
\begin{center}
$\begin{array}{c}
\includegraphics[width=8cm]{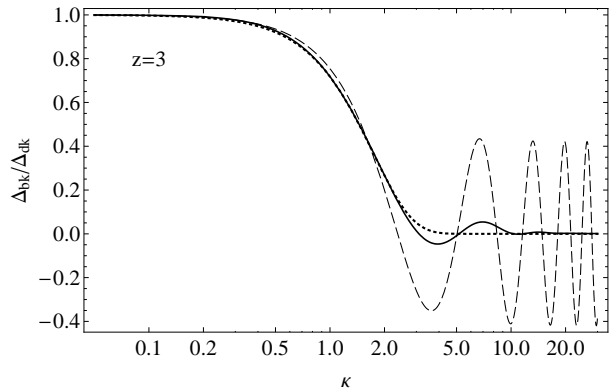}
\end{array}$
\end{center}
\caption{ The same plot as Fig. \ref{solo} but has the dotted curve represent the fitting function $\Delta_{bk}/\Delta_{dk}=\exp(-k^2/k_f^{\prime2})$ with the modified filtering wavenumber $k_f^{\prime}$.
}\label{sol3}
\end{figure}
\section{Results}
\label{res}
\begin{figure}[tb]
\begin{center}
$\begin{array}{c}
\includegraphics[width=8cm]{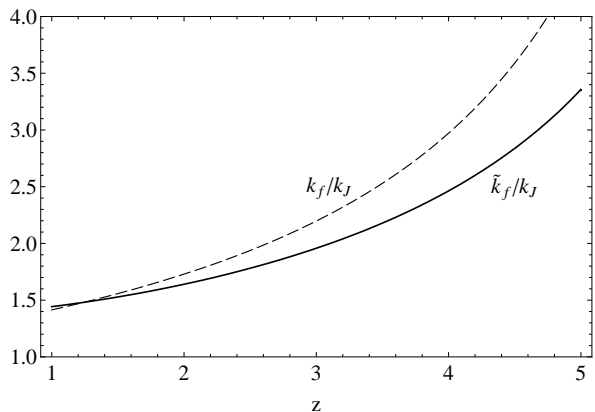}
\end{array}$
\end{center}
\caption{ The comparison of two characteristic scales in gas and plasma approaches versus  redshift $z$. The solid curve is $k_f^{\prime}/k_J$ and the dashed curve is $k_f/k_J$ where $k_J$ is the ordinary Jeans scale for baryon gas.
}\label{filtering}
\end{figure}
\begin{figure}[tb]
\begin{center}
$\begin{array}{c}
\includegraphics[width=8cm]{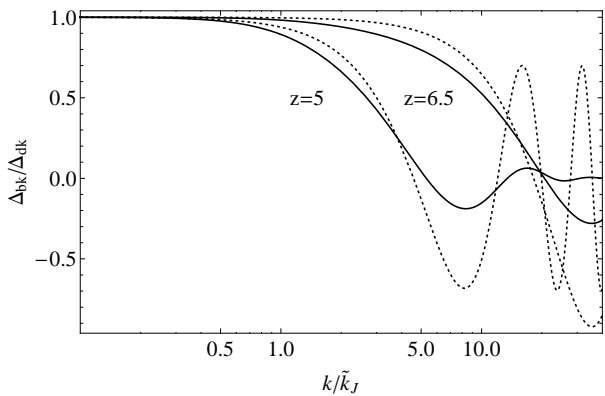}
\end{array}$
\end{center}
\caption{ The comparison of baryon density spectra at $z=6.5$ and $z=5$ for gas (dashed) and plasma (solid) approaches.
}\label{early}
\end{figure}
 With the Landau damping introduced, we now calculate the complete baryon density spectrum by combining Eq.~(\ref{sol}) and~(\ref{baryld}). Figure \ref{solo} shows the result (in solid curve) at $z=3$ with $z_{ion}=7$. The dashed and dotted curves correspond to the results obtained from the neutral gas hydrodynamics and plasma physics without applying the Landau damping. We see that the dotted curve shifts from the dashed one by $\sim 10\%$ due to the increase of the Jeans wavelength mentioned in Sec. \ref{ion}. When suffering the Landau damping, the modes of baryon overdensity at large $k$ diminish, leading to a suppression of the baryon density spectrum in the intermediate regime of $k$. Such a suppression, however, accidentally compensates the shift of the spectrum. Thus the filtering wavenumber $k_f$ proposed by Gnedin and Hui \cite{filtering} to characterize the baryon structure would not be suitable when the Landau damping effect is included.

In an expanding universe, the Jeans scale changes with temperature and therefore with time. On the other hand, it takes about one Hubble time for baryons to respond to the changed pressure. Therefore the filtering wavenumber $k_f$ was introduced \cite{filtering,sound:Gnedin} to account for the migration of the Jeans wavenumber over the time scale relevant to the acoustic oscillations, which therefore depends on the entire previous thermal history of the universe \cite{filtering2}. In the matter-dominant era $k_f$ is related to  $k_J$ by \cite{sound:Gnedin}
 \begin{equation}
    \frac{1}{k_f^2}=\frac{3}{a}\int_{0}^{a}\frac{da^{\prime}}
 {k_J^2(a^{\prime})} 
 \left[1-\left(\frac{a^{\prime}}{a}\right)^{1/2}\right].
 \end{equation}
When the baryon fluctuations suffer the Landau damping effect, a correction to $k_f$ must be taken. The original $k_f$ was deduced by supposing the solution $\Delta_{bk}/\Delta_{dk} = 1-k^2/k_f^2\approx \exp(-k^2/k_f)$ under the small $k$ approximation ($k \to 0$). With the exponential damping factor included, we find it reasonable to directly assume $\Delta_{bk}/\Delta_{dk}(\alpha,\kappa) \approx \exp(-k^2/\tilde{k}_f^2)$, where $\tilde{k}_f$ is the modified filtering wavenumber. Assuming $\tilde{k}_f/k_J= \sigma$, we have 
\begin{equation}
 \exp(-\frac{\kappa^2}{\sigma^2})=e^{-0.4(q/4)\ln\alpha}b_i(\alpha,\kappa)+\displaystyle\frac{1}{1+\kappa^2}.
\end{equation}
Upon taking logarithm and then square-root on both sides, we obtain
\begin{equation}
  \frac{1}{\sigma}=-\frac{1}{\kappa}\sqrt{\ln \left(e^{-0.1q\ln\alpha}b_i(\alpha,\kappa)+\frac{1}{1+\kappa^2}\right)}.
\end{equation}
The right-hand side of the above expression approaches a constant for $\kappa\in[0.8,2]$ at  $z<5$, which falls in the intermediate regime of $k$. Hence the modified filtering wavenumber $\tilde{k}_f$ can be obtained by substituting an arbitrary $\kappa$ within this range. Figure \ref{sol3} is the same plot as Fig.\ref{solo} but with an additional dotted curve that illustrates the well-fitted function of $\exp(-k^2/\tilde{k}_f^2)$. In Fig.~\ref{filtering} we compare the two characteristic scales $k_f$ (dashed) and $k_s$ (solid) at different redshifts. The two curves converge to each other at low redshifts, while in general $\tilde{k}_f$ is smaller than $k_f$ until $z \sim 1$.  
 
\section{Conclusion}
  In summary, the baryon overdensities has three stages of evolution: 1) before the photon-baryon decoupling, where baryons were in a plasma state but were tightly coupled to and dominated by the photons; 2) during the dark age, where baryons formed neutral gases and grew with dark matter overdensities; and 3) after reionization, where baryons turned into a plasma again and followed the plasma hydrodynamics. Although before the decoupling baryons were in a plasma state, the photon-baryon fluid was dominated by the photon pressure rather than the combination of the baryon thermal and electrostatic pressures. This resulted in a sound speed that is much larger than the ion thermal velocity. The plasma effects were negligible in that epoch. This aspect has been widely recognized.
  
What appears less recognized is the influence of the plasma collective effects in the post-reionization epoch. Considering the plasma effects during this epoch, we point out that the baryons would experience two major impacts: the change of the sound speed and the Landau damping of the baryon oscillations. Numerically speaking, the first effect is relatively minor compared with the second. Since $T_i \approx T_e$, we expect that the Landau damping would be quite effective in dissipating the amplitude of baryon oscillation modes. This might influence the subsequent galaxy formations history.
According to Eq.~(\ref{baryq}), the amplitude of baryon oscillations scales as $\alpha^{-5/4}$. Therefore the discrepancy between the gas and the plasma approaches is most significant in the early stage of structure formation immediately after the reionization. Figure~\ref{early} is a comparison of the baryon density spectra based on the gas (in dashed line) and the plasma (in solid line) approaches at different redshifts ($z=6.5$ and $z=5$) with $z_{ion}=7$. If the baryons behave as a neutral gas, then we expect a sizable signature of baryon acoustic oscillations in the matter power spectrum below the scale $k> O(1)h{\rm Mpc}^{-1}$ after the photo-reheating process. Conversely, if the baryons act more like a plasma, then we should find little sign of oscillations but instead the Landau damping of it, whose rate maybe measured via tracing the spectrum back to the early time. So far the observations of the matter power spectrum have not yet reached the sensitivity for such small scales at high redshifts \cite{Tegmark}. It would be interesting if this regime can be explored in the near future, so that we can learn more about the nature of the cosmic structure formation.

\section{Acknowledgment}
This research is supported by Taiwan National Science Council under Project No. NSC 97-2112-M-002-026-MY3, by Taiwan's National Center for Theoretical Sciences (NCTS), and by the US Department of Energy under Contract No. DE-AC03-76SF00515.

\end{document}